\documentstyle[pre,aps,epsfig]{revtex}
\tighten
\draft

\def\be{\begin{equation}} 
\def\bea{\begin{eqnarray}} 
\def\ee{\end{equation}}
\def\eea{\end{eqnarray}}

\def\iff{\infty}
\def\amax{(\alpha_{max}-\alpha_{min})}
\def\amaxf{(\alpha-f(\alpha))}
\def\dal{\Delta_{\alpha}}

\begin{document}
\twocolumn[\hsize\textwidth\columnwidth\hsize\csname @twocolumnfalse\endcsname
\title{ Electron delocalization and multifractal 
scaling in electrified random chains}
\author{Parthapratim Biswas\cite{PARTHA} and
Prabhat K Thakur\cite{PRABHAT}
}
\address{
Condensed Matter Theory Group\\
S.~N.~Bose National For Basic Sciences\\
Salt Lake City, Block JD, Sector III\\
Calcutta, 700 091 INDIA\\
}
\maketitle
\begin{abstract}

Electron localization property of a random chain changing under 
the influence of a constant electric field has been studied. We 
have adopted the multifractal scaling formalism to explore the 
possible localization behavior in the system. We observe 
that the possible localization behavior with the increase of electric 
field is not systematic and shows strong instabilities associated 
with the local probability variation over the length of the chain.
The multifractal scaling study captures the localization aspects 
along with strong instability when the electric field is changed 
by infinitesimal steps for a reasonably large system size. \\ \\

 
\end{abstract}
\vskip 1.0cm
]

Electronic states are exponentially localized in one 
dimensional (hereafter $1D$) random chains and the envelope 
of the wave function, $\phi(x) \sim 
\exp{(-\alpha \, x)}$ for $x \to \iff$ where $\alpha$ is the 
inverse of localization length \cite{kn:Mott,kn:Borland}. 
This localization nature of electronic states can be changed 
through application of a constant electric field and as a result 
electronic states exhibit some kind of different localized 
nature over the sample size. The problem of electronic states 
and localization in a random chain in the presence of a constant 
electric field is still a matter of controversy and is yet to be 
fully understood \cite{kn:Ben,kn:Cota1,kn:Sokol,kn:Delyon,kn:Cota2,kn:Gil}. 
In the past, the possibility of existence of 
non-exponential localization or localization/delocalization 
transition in electrified chain has been addressed \cite{kn:Sokol,kn:Delyon}. 
The deviation from exponential localization has been also claimed through 
the numerical study of electronic transmittance \cite{kn:Sokol}.

However, the localization mechanism can be understood more rigorously  
within the multifractal scaling analysis of the electronic wave 
functions without any {\em a priori} assumption of exponential 
localization nature and the existence of localization length 
\cite{kn:Thakur1,kn:Schreiber,kn:Thakur2,kn:Thakur3}.
The multifractal scaling formalism \cite{kn:Chabbra} has been 
invoked in the recent past to analyze the nature of electronic 
states in the vicinity of the mobility edge \cite{kn:Schreiber,kn:Thakur2}, 
and also for characterization of critical nature of electronic states 
in $1D$ Fibonacci quasiperiodic systems \cite{kn:koho1,kn:koho2}. 
Generally speaking, in all of the above examples, wave functions 
exhibit a rather involved oscillatory behavior displaying strong 
fluctuations. As a consequence the notion of envelope 
wave function or lyapunov exponent which has been successful 
in studying both the extended and exponentially localized states, is 
no longer suitable to the states in examples above. 
On the other hand multifractal scaling formalism has been found to be 
very useful to characterize the spatial fluctuating pattern 
of the wave functions which are neither Bloch like homogeneously 
extended state nor the exponentially decaying one. The  
same scaling analysis has been successful to extrapolate in the 
extreme limits also. So, it naturally appears that the possible 
delocalization behavior that we are going to address in this 
Letter can also be understood through the same scaling analysis. 

The aim of this Letter is to report our investigation on 
how does localization nature is influenced by switching on a constant 
electric field through numerical study of electronic wave functions 
for a reasonably large array of $\delta$-function random potentials 
having a bi-modal distribution. The choice of this type of potential 
has been justified in the past both from its experimental relevances 
as well as from the pure academic interest. We start with the 
Schr\"odinger's equation for the electrons in a random chain in 
presence of a constant electric field : 
\be
\left[-\frac{d^2}{dx^2}+\sum_{n=1}^{N} V_{n}\delta(x-na)-Fx \label{eqn-sch1}
\right]\Psi(x) = E \Psi(x)
\ee
where the units are such that $(\hbar^2=2m_e=1)$
$m_e$ is the effective mass of electron. The electric field induced 
force $F$ is expressed in unit of $\left(\frac{\hbar^2}{2m_e a^3}\right)$, $a$
is the lattice spacing and $V_{n}$ is the strength of the $n$-th
potential barrier (taking the value $V_A$ or $V_B$ randomly), 
$F$ is the product of electric field by the electronic charge.
The lattice constant $a$ is taken unity throughout this calculation.

One can map the above Eq.(\ref{eqn-sch1}) to a  finite difference 
equation by approximating the potential $Fx$ by a step function in-between 
the $\delta$-functions. Within this approximation, the solutions in-between 
the $\delta$-function potentials are now plane waves 
instead of Airy functions. The corresponding Poincare map 
\cite{kn:Sokol,kn:Macia,kn:Bell} is: 
\be
\Psi_{n+1}=A_n\Psi_{n}+B_n\Psi_{n-1}
\ee
The coefficients $A_n$ and $B_n$ are given by \cite{kn:Sokol}
\bea
A_n &=& \left[\cos k_{n+1}+\frac{k_n}{k_{n+1}}\frac{\sin k_{n+1}}
{\sin k_n}\cos k_n+V_n \frac{\sin k_{n+1}}{k_{n+1}}\right] \nonumber \\
B_n &=& -\frac{k_n} {k_{n+1}}\left(\frac{\sin k_{n+1}}{\sin k_n}\right)
\eea
with $k_n= (E+n\,F)^{1/2}$ and $\Psi_n= \Psi(x=n)$. 
Now in order to solve the equation iteratively for a reasonably 
large system size one can consider the initial values for 
$\Psi_1 = \exp(-\imath E^{1/2}a)$ and $\Psi_2 = (-2\imath E^{1/2}a)$, $E$ 
being the incident electron energy before it reaches the region 
where the electric field is applied. 

The transmittance corresponding to the array of random 
$\delta$-function potential is given by \cite{kn:Sokol}
\be
T=\left(\frac{k}{k_1}\right)
\left(\frac{|\exp (2\imath k_1)-1|^2}
{|\Psi_{N+2}-\Psi_{N+3}\exp (-\imath k_1)|^2}\right)
\ee
with
$$k=E^{1/2}\quad \mbox{and} \quad k_1= (E+F L)^{1/2}$$ and $L=Na$.

We now analyze the pattern of local probability 
density $|\Psi_n|^2 $ along the chain through its
multifractal scaling relation of $\alpha$ and $f(\alpha)$, where $\alpha$ 
stands for the scaling exponent and $f(\alpha)$, the corresponding 
distribution function. We have used the mathematical prescription 
suggested by Chabbra and Jensen \cite{kn:Chabbra} for its 
simplicity and success in correctly evaluating the quantities $\alpha$ 
and $f(\alpha)$ directly through the normalized measure without any numerical 
instability. Let us define the required normalized measure in our study by 
$$ 
P_i = \frac{|\Psi_i|^2}{\sum_{i=1}^N|\Psi_i|^2}
$$
where the scaling behavior of $P_i \sim N^{-\alpha_i} $ 
for $N \to \iff$. 

According to Chabbra and Jensen if we define the  $q$-th moment 
of the probability measure $P_i$ by $\mu_{i}(q,N)$ where
$$
\mu_i(q,N)= \frac{P_i^q}{\sum_{i=1}^N P_i^q}
$$ 
then a complete characterization of the fractal singularities can 
be made in terms of $\mu_i(q,N)$. The  expression for the distribution 
function of scaling exponent $\alpha$ can be written as:

\be
f(\alpha) = \lim_{N \to \iff} -\frac{1}{\log N}
\sum_{i=1}^{N}\mu_{i}(q,N)\,\log \,\mu_i(q,N)
\ee 

and the corresponding singularity strength of the measure 
is obtained by 
\be
\alpha = \lim_{N \to \iff}  -\frac{1}{\log N}
\sum_{i=1}^{N} \mu_{i}\, \log\, P_i. 
\ee

One can infer on the nature of electronic states for large $N$ 
based on the following observation:

\begin{enumerate}

\item 

{\em Extended nature} : $\alpha_{min} \to 1 $, $f(\alpha_{min}) \to 1 $, 
                    $\alpha_{max} \to 1 $, $f(\alpha_{max}) \to 1 $.

\item 
{\em Localized nature} : $\alpha_{min} \to 0 $, $f(\alpha_{min}) \to 0 $, 
                    $\alpha_{max} \to \iff $, $f(\alpha_{max}) \to 1$.

This property is usually manifested by rectangular 
two-hump form of $f(\alpha)$ curve with a sparse distribution 
of points in between.

\item 
{\em Critical nature} : $\alpha $ vs $f(\alpha)$ curves closely overlap  
                  on one another with the increase of system size.

\item 
{\em Power-law nature} : The right portion of $\alpha-f(\alpha)$ 
curve deviates slowly from one another with the increase of 
system size in contrast to the strong deviation as seen in 
the exponential decay.
\end{enumerate}

We now define a simple way the degree of localization.
We consider $\dal=\amax$ as a measure of degree of 
localization. This can distinguish clearly an extended state 
from a localized state for a sufficiently large system size $N$.
Also, one can investigate the change in the nature of states brought 
about due to the change of some external parameters, e.~g., electric 
field. 

In figure (1) we have shown the spatial pattern of local probability 
variation along the chain for a 
localized state in both zero and finite electric field. The upper 
curve exhibits the delocalized pattern in the presence of electric 
field $F=1.25 \times 10^{-5}$ unit. The corresponding localized and 
delocalized  behavior of the electronic transmittance data have been 
been presented for both the zero as well as for a finite electric 
field in figure (2). In figure (3), we have shown $\amaxf$ 
plots for both the zero and the finite field value $2\times 10^{-5}$. 
In the zero field case 
the plot shows in $f(\alpha)$ a two-hump form corresponding to the
exponential localization. This is due to the fact that $\amaxf$ spectrum 
is densely populated on the extreme left and on the right 
but the region in-between is very sparse. On the other hand, for
finite electric field, the $\amaxf$ data gather in a relatively 
narrow region on the curve having a convex shape. 
This indicates that the change in localization behavior shows 
an overall delocalization trend which is exhibited through the more 
spatial extension of the state over the sample and hence resulting 
in the contraction of the $\alpha-f(\alpha)$ spectrum. 

Next we investigate further whether this delocalization pattern can change 
systematically as we change the electric field through small steps 
of the order of $10^{-8}$ unit for a sufficiently large system size.

In figure (4) we have shown the plot of degree of localization 
$\dal$ with the electric field for two large system
sizes, $N=10^5$ (upper curve) and $N=2\times 10^5$ (lower curve). 
In both the plots we see that $\dal$ exhibits strong instability 
throughout the whole regime of electric field from a very low to 
as high as $10^{-5}$ unit. We have observed that for both the two 
system sizes, the order of deviation in $\dal$ from its previous 
step is of the order of unity whereas the value of $\dal$ itself is 
of the same order. Also the fluctuating pattern of $\dal$ with $F$ 
for the both the two system sizes is almost the same.
In figure (5) we have shown again the variation of the degree
of localization, i.~e., $\dal$ with the electric field for a 
different set (cf. Figure 4) of potential parameters. Here the 
order of fluctuations in $\dal$ corresponding to two large 
systems, i.~e., $75\times 10^3$ and $1.7\times 10^4$ number of 
atoms have been presented. The order of fluctuations appear to be 
quite significant and it is nearly unity in the both the cases over 
a wide region of the electric field as shown in the figure through 
covering the fluctuating zone between the horizontal lines.

We think this kind of instability has its intrinsic origin in the 
restructuring of the different states in a complicated manner and 
each of them is highly sensitive even for an infinitesimal change 
in the applied electric field. This can be also understood as due to 
the competing nature of the potential $Fn$ and the disordered 
$\delta$-potentials in the large system size. However, if one neglects 
this fluctuation through some brute force methods, it shows only an 
apparent simple delocalization effects due to the increase of electric 
field. 

In conclusion, we have shown that the change of localization 
aspects due to the increase of a uniform electric field is not 
simply an overall delocalization but the localization property 
is very sensitive with respect to an infinitesimal change of 
electric field giving rise to a strong instability in the degree 
of localization. This instability aspect in the localization/delocalization 
is present for all reasonably large lengths for an appropriate 
set of parameters and is due to the combined effects of disorder 
potential and the electric field induced linear potential. At 
sufficiently large length scale the states change depending upon 
the restructuring of the spectrum from its previous form and hence 
the localization property of a particular state in a given field will 
change drastically giving rise to a state of modified localized nature.
P.~Biswas would like to thank the Council of Scientific and 
Industrial Research (CSIR) for financial assistance in the form of a 
senior research fellowship.

\begin{figure}
\label{fig1}
\epsfxsize=3.0in \epsfysize=3.0in
\rotatebox{270}{\epsfbox{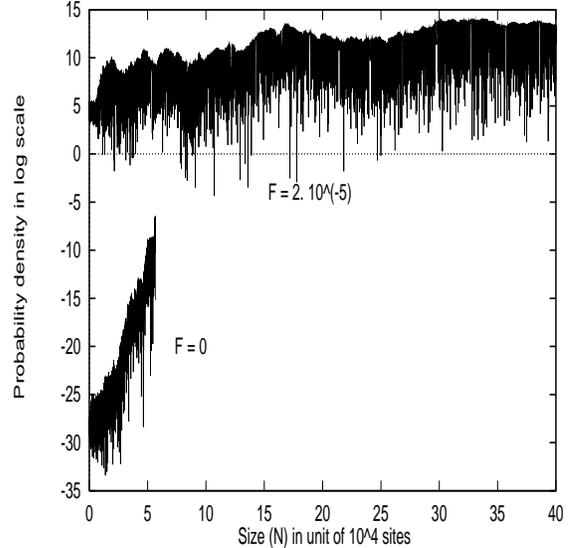}}
\caption{
Local probability (in log-scale) vs system size ($N$) in unit of $10^4$ number 
of sites. The parameters are : $V_A=1.25$ units, $V_B=1.4$ units, $c=0.01$, 
$N=4\times 10^{5}$ for electric field $F=2.0\times 10^{-5}$ (upper) unit and 
$N=5.6\times 10^4$ for zero field (lower) with the energy of the incoming 
electron $E=1.4890$ units in both cases. The data in the upper curve has been 
given a constant shift of $20$ units in the log-scale for convenience of 
comparison.}
\end{figure}

\begin{figure}
\label{fig2}
\epsfxsize=3.0in \epsfysize=3.0in
\rotatebox{270}{\epsfbox{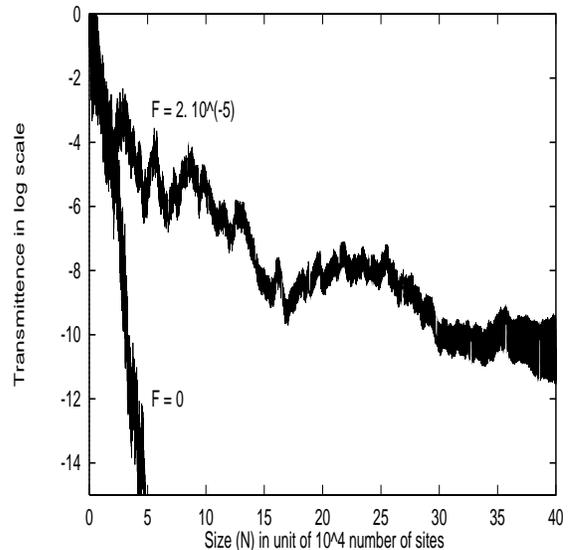}}
\caption{
Transmittance $T$ (in log-scale) vs size ($N$) for the set of parameters 
as in figure 1. The plot is made up to a length of $4\times 10^{5}$ number 
of sites for $F = 2.0\times 10^{-5}$ (upper curve) and up to $5\times 10^4$ 
number of sites in the zero field case (lower curve).
}
\end{figure}

\begin{figure} 
\label{fig3}
\epsfxsize=3.0in \epsfysize=3.0in
\rotatebox{270}{\epsfbox{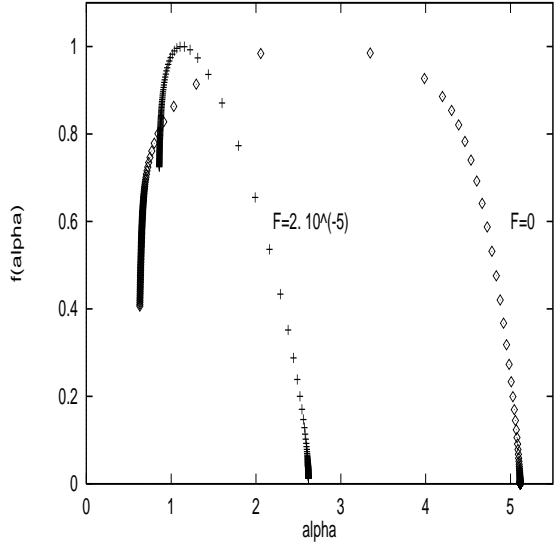}}
\caption{ 
$\alpha$ vs $f(\alpha)$ plot for finite ($F=2\times 10^{-5}$ unit) as well 
as zero electric field for the local probability 
variation ($|\Psi_n|^2$) over the chain of length $N=4\times 10^5$. The 
other parameters are the same as in figure (1).
}
\end{figure}

\begin{figure}
\label{fig4}
\epsfxsize=3.0in \epsfysize=3.0in
\rotatebox{270}{\epsfbox{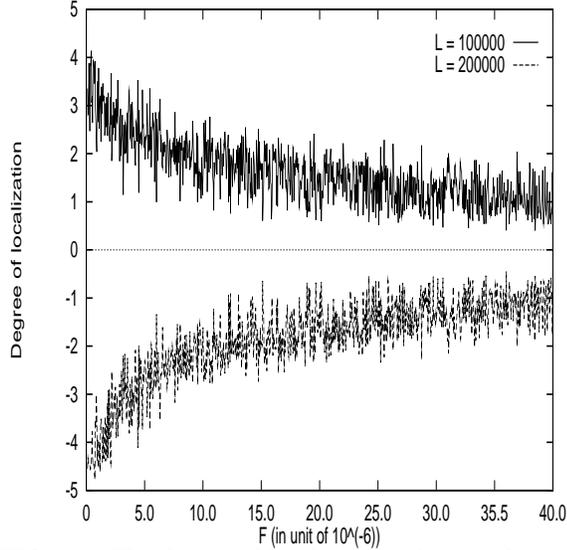}}
\caption
{ 
The degree of localization ($\dal$) vs electric field ($F$) for the two 
different system sizes $10^5$ (thick line) and $2\times 10^5$ (dashed line).
The parameters are $V_A = 1.25$ units, $V_B = 1.4$ units, $c=0.01$, and 
for incoming energy $E= 1.4890$ units. Once again, for comparison, we plot 
the negative of $\dal$ for system size $2\times 10^5$ (dashed line).
}
\end{figure} 

\begin{figure}
\label{fig5}
\epsfxsize=3.0in \epsfysize=3.0in
\rotatebox{270}{\epsfbox{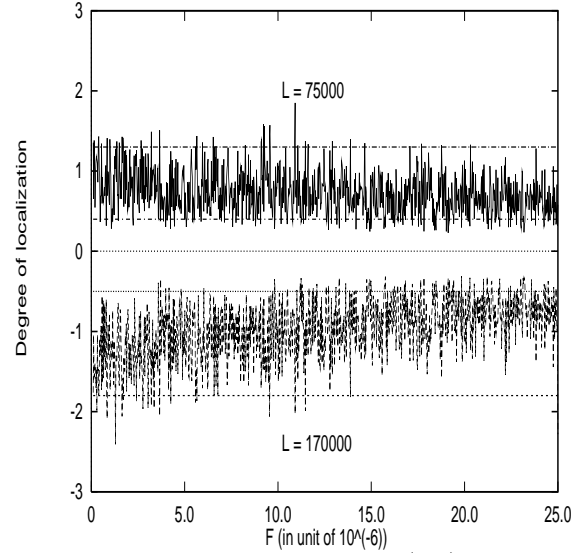}}
\caption{ 
The degree of localization ($\dal$) vs electric field ($F$) for a 
different set of potential parameters. The thick and the dashed line 
corresponds to system size $7.5\times 10^4$ and $1.7\times 10^5$  
with other parameters, $V_A = 5.15$ units, $V_B = 5.34$ 
units, and $c=0.05$, and energy $E=5.6025$ units.
}

\end{figure}

\end{document}